\documentclass[%
 reprint,
showpacs,preprintnumbers,
 amsmath,amssymb,
 aps,
]{revtex4-1}

\usepackage{graphicx}
\usepackage{dcolumn}
\usepackage{bm}

\begin{document}

\title{Forces between a partially coherent fluctuating source and a magnetodielectric particle. }

\author{Juan Miguel Aun\'{o}n }
\author{Manuel Nieto-Vesperinas}
\email{mnieto@icmm.csic.es}

\affiliation{Instituto de Ciencia de Materiales de Madrid, Consejo Superior de
Investigaciones Cient\'{i}ficas\\
 Campus de Cantoblanco, Madrid 28049, Spain}
 
\begin{abstract}
We address the  forces exerted by the electromagnetic field emitted by a planar fluctuating source on dielectric particles that have  arose much interest because of their recently shown magnetodielectric behavior. In this context, we analyze as a particular case the modification of the Casimir and Van der Waals forces. We study the effect of the source coherence length as well as the interplay between the force from the radiated field and that from the electric and magnetic dipoles induced on the particle. This allows a control of these interactions as well as of the weight and interference effects between the fields from both kinds of induced dipoles, in particular when large changes in their differential scattering cross section occur due to Kerker minimum forward or zero backward conditions; thus opening new paths to nanoparticle ensembling and manipulation. The influence of surface waves of the source is also studied.
\end{abstract}



\maketitle

Magnetodielectric particles have recently attracted much attention due to their exotic properties as scatterers and nanoantennas \citep{novotny2011antennas,mehta2006experimental, Liu2012Broadband,schuller2010plasmonics,Rolly2012boostingthedirectivity, Schuller2007Dielectricmetamaterials,Kerker83Electromagnetic,kuznetsov2012magnetic,Evlyukhin2012MagneticDipole}. It has been shown \citep{GarciaEtxarri11Anistropic,Lukyanchuk2010hugelight} that some dielectric particles behave in this way exhibiting coupled electric and magnetic dipoles induced by the illuminating light field. It is then of great interest to study their response to stochastic radiation forces, specifically those due the field of a planar fluctuating source \citep{mandel1995optical}. Many previous studies have dealt with this subject for atoms or non-magnetic nanoparticles with its application to delta-correlated thermal sources and blackbodies in connection with the Van der Waals (VdW) and Casimir-Polder (C-P) interactions \citep{Henkel2002Radiationforces,Antezza2005NewAsymptotic, Novotny2008vdw}. 

In this work we deal with a more general kind of statistical sources, namely those that are spatially partially coherent and of the wide variety of those statistically homogeneous and isotropic \citep{mandel1995optical,aunon2012opticalforces}. 
 Their emission excites electric and magnetic dipoles of the particle in its near field that may be considered as a secondary source whose radiation interacts with the primary fluctuating source; this giving rise to a new total force resulting from both the action of the field radiated by the primary source and that from the field emitted by these dipoles. At thermal wavelengths, the interaction from this secondary source constituted by the particle induced dipoles is interpreted as a Liftshitz force \citep{lifshitz1956theory}, which   at zero  temperature becomes either  those derived by VdW and  C-P \citep{casimir1948}, depending on the distance and use, or not, of quasistatic formulations. However, in our study, first the optical wavelengths are such that $\hbar\omega/kT>>1$ and hence Planck energy becomes similar to that of the vacuum fluctuations: $\hbar\omega[\frac{1}{2}+1/(\exp(\hbar\omega/kT)-1)]\approx  \frac{1}{2}\hbar\omega$; and second, due to the magnetic response of the nanoparticle, more forces come into play in addition to those that keep an analogy with these above quoted, thus allowing a larger number of degrees of freedom of relevance for particle ensembling and manipulation. We do not restrict here to  thermal wavelengthts, but rather consider  a range of infrared and optical frequencies whose choice depends on the particle size. Then a particular case of our study is that analogous to those dipole forces    induced by spontaneous electromagnetic field fluctuations. While the vacuum forces can become relatively  negligible by conveniently manipulating the power intensity of the source, we shall show that in other possible configurations and experimental designs they  may predominate.

The geometry considered in this letter consists of two half-spaces. The lower one ($z<0$) is occupied by the  source with its polarization currents and it will be denoted as $1$; whereas the upper one  ($z>0$), denoted as $2$,  is free space and contains the particle. 


The Cartesian components of the force on a  magnetodielectric dipolar particle is the sum of an electric, magnetic and electric-magnetic dipole interference parts which are  expressed in terms of the first electric and magnetic Mie coefficients $a_{1}$  and $b_{1}$ as \citep{nieto2010optical}
\begin{eqnarray}
F_{i}\left(\mathbf{r}\right)	&=&	F_{i}^{e}\left(\mathbf{r}\right)+F_{i}^{m}\left(\mathbf{r}\right)+F_{i}^{e-m}\left(\mathbf{r}\right) \nonumber \\
	&=&\frac{\varepsilon_{0}\varepsilon_{2}}{2}\text{Re}\left\{ \left\langle \alpha_{e}E_{j}^{*}\left(\mathbf{r}\right)\partial_{i}E_{j}\left(\mathbf{r}\right)\right\rangle \right\} \nonumber \\
	&+&	\frac{\mu_{0}\mu_{2}}{2}\text{Re}\left\{ \left\langle \alpha_{e}H_{j}^{*}\left(\mathbf{r}\right)\partial_{i}H_{j}\left(\mathbf{r}\right)\right\rangle \right\} \nonumber \\
	&-&	\varepsilon_{0}\varepsilon_{2}\frac{Zk_{0}^{4}}{12\pi}\text{Re}\left\{ \left(\alpha_{e}^{*}\alpha_{m}\right)\left\langle \mathbf{E}^{*}\times\mathbf{H}\right\rangle _{i}\right\} ,
\end{eqnarray}
where $i,j=1,2,3$, $\varepsilon_l=\varepsilon_l'+i\varepsilon_l''$ and $\mu_l=\mu_l'+i\mu_l''$ $(l=1,2)$ are the permittivity and  susceptibility of the medium embedding the particle, respectively,  in our case being vacuum; and $Z=\sqrt{\mu_0\mu_1/\varepsilon_0\varepsilon_1}$. $\alpha_e$ and  $\alpha_m$ standing for the electric and magnetic polarizability of the particle, respectively. $
 \alpha_e =  i \frac{3 \epsilon_0 } {2 k ^3} a_1$, $\alpha_m = i \frac{3} {2\mu_0 k ^3} b_1$; $E_{i}(\bf{r})$ is the total electric vector at  frequency $\omega$ at  any point of the half-space $z>0$, hence at the position of the particle, i.e. at $\mathbf{r}=\mathbf{r}_{0}$, it will be
\begin{eqnarray}
E_{i}(\mathbf{r}_{0})&=&E_{i}^{inc}(\mathbf{r}_{0})+E_{i}^{p}(\mathbf{r}_{0})+E_{i}^{m}(\mathbf{r}_{0})\nonumber \\
&=&E_{i}^{inc}+\mu_{0}\mu_{2}\omega^{2}G_{ij}^{p}p_{j} +i\mu_{0}\mu_{2}\omega G_{ij}^{m}m_{j} 
\label{totfield}.
\end{eqnarray}
 In the second line of this equation we have omitted the explicit dependency on space and frequency for brevity. The associated magnetic field can be obtained directly from Maxwell's equations.  In Eq. (\ref{totfield}) the last two terms are the electric fields emitted by the particle induced dipoles ($p_i=\varepsilon_0\alpha_eE_i^{inc}$ and $m_i=\alpha_mH_i^{inc}$) after multiple reflections   at the plane $z=0$ \citep{sipe1987quantum}  and hence connect the constitutive properties of the source and particle through the reflection Fresnel coefficients $r_{s,p}$  at $z=0$ and the polarizabilities, being described by $G_{ij}^{p,m }$. On the other hand,  the first term $E_{i}^{inc}$ represents the electric field \textit{incident} on the particle after being emitted from the primary source placed at $z<0$, and is defined through the Green's function $G_{ij}^{P}$ which includes the transmission Fresnel coefficients $t_{s,p}$  from $z<0$ into $z>0$. $G_{ij}^{P}$ may be written as a superposition of plane waves  with  wavevector 
$k_0\mathbf{s}_{i}=k_0(\mathbf{s_{\perp}},s_{z,i})$ with  $\mathbf{s_{\perp}}=(s_x, s_y)$, $s_{z,i}^{2}=\varepsilon_{i}\mu_{i}-s_\perp^2$,($i=1,2$), and $k_0=2\pi/\lambda=\omega/c$. Thus, in terms of the polarization currents one has 
\begin{equation}
E_{i}^{inc}\left(\mathbf{r}_{0}\right)	=	\mu_{0}\mu_{2}\omega^{2}\int_{V}G^{P}_{ij}\left(\mathbf{r}_{0},\mathbf{r}',\omega\right)P_{j}\left(\mathbf{r}',\omega\right)d^{3}r' .
\end{equation}
In vacuum ($\varepsilon_{2}=\mu_{2}=1$), $G_{ij}^{p,m}$ decays exponentially with the distance in the evanescent wave region ($s_\perp>1$) and  is oscillatory in the radiative one ($s_\perp<1$)\citep{novotny2006principles}. For a single dipole, this electric field is calculated from Eq. (3) using $G_{ij}^{p}$ and $p_i\delta({\bf r}'-{\bf r}_0)$ instead of $G_{ij}^{P}$   and $P_{j}({\bf r}')$. In a similar way,  the electric field $E^m$ generated by the magnetic dipole  can be calculated from the Green function associated to the magnetic field of the electric dipole $H^p$  and making the interchange $r^{s}\leftrightarrow r^{p}$ in the reflection Fresnel coefficients. Some details about these Green's functions can be found in e.g.  \citep{Sipe1987green,Joulain2005Surfaceelectromagnetic}. Note that for p-polarization, both  $t_p$ and $r_p$ support surface plasmon polaritons (SPPs) when $\varepsilon_2'<1$. 

Now we define the cross-spectral density tensor  of the source polarization as $W_{ij}^{(P)}\left(\mathbf{r}_{1},\mathbf{r}_{2},\omega\right)=\left\langle P_{i}^{*}\left(\mathbf{r}{}_{1},\omega\right)P_{j}\left(\mathbf{r}{}_{2},\omega\right)\right\rangle $. We shall address the wide variety of non-local statistically homogeneous and isotropic sources  \citep{mandel1995optical}) for which 
\begin{equation}
W_{ij}^{(P)}(\mathbf{r}_{1},\mathbf{r}_{2},\omega)=S^{(P)}(\omega)\mu_{ij}^{(P)}(|\mathbf{r}_{1}-\mathbf{r}_{2}|,\omega).
\end{equation}
$S^{P}(\omega)$ denotes the power spectrum of the source and $\mu_{ij}(\mathbf{r}_{1},\mathbf{r}_{2},\omega)$ is the spectral degree of coherence \citep{mandel1995optical}. A special particular case of these sources are those thermal and blackbodies  widely studied.
On inserting Eq. (2) into (1) and taking the statistical homogeneity of the source into account, one obtains the total force on the particle. For these sources only the force along the $z$-axis is different from zero. We consider mutual incoherence between the particle electric and magnetic induced dipoles, i.e., $\left\langle p_i^*m_j \right\rangle=0$ \citep{radiophysics}. $F_{1}$ and $F_{2}$ will denote  the total forces due to the above mentioned contributions of the primary fluctuating source in $z<0$  and to the secondary source field from the particle induced dipoles, respectively.

We study the effect of  the magnetodielectric properties of the particle in the near infrared. This is a semiconductor sphere; its anomalous scattering properties have recently received a great deal of attention, both theoretically and experimentally \citep{nietoJOSA011,nieto2012nature, Lukyanchuk2010hugelight, Liu2012Broadband}. In particular, for each incident plane wave component its scattered intensity in the backscattering direction is  zero, [first \textit{Kerker condition} ($K1$)] when $\text{Re}{\alpha_e}=\text{Re}{\alpha_m}$ is fulfilled. Also for each of such plane wave components impinging the particle, the forwardly scattered intensity  becomes close to  a non-zero minimum [second \textit{Kerker condition}  (K2)]  when $\text{Re}{\alpha_e}=-\text{Re}{\alpha_m}$ . In both cases: $\text{Im}{\alpha_e}=\text{Im}{\alpha_m}$ \citep{nietoJOSA011}. 

Thus, we address a  Si sphere of radius  $a=230 nm$, the spectrum of the incident light being  in the range of $1.2-2 \mu m$. At these frequencies the total cross section of the particle is  fully determined by the Mie coefficients $a_1$ and $b_1$ \citep{GarciaEtxarri11Anistropic}, this justifies the use of Eq. (1).

We assume a Gaussian degree of coherence of the source, therefore the correlation function reads $W_{ij}^{(P)}\left(\mathbf{r}_{1},\mathbf{r}_{2},\omega\right)$ $={\cal S}^{(P)}\left(\omega\right)\text{exp}\left(-\left(\left|\mathbf{r}_{1}-\mathbf{r}_{2}\right|\right)^{2}/2\sigma^{2}\right)\delta_{ij}/(2\pi)^{3/2}\sigma^{3}$, $\sigma$ being  the coherence length of the source and $S^{(P)}(\omega)={\cal S}^{(P)}(\omega)/(2\pi)^{3/2}\sigma^{3}$ representing the normalized spectrum. Notice that for $\sigma\rightarrow 0$, $W^{(P)}_{ij}(\mathbf{r}_1,\mathbf{r}_2,\omega)={\cal S}^{(P)}(\omega)\delta(|\mathbf{r}_1-\mathbf{r}_2|,\omega)\delta_{ij}$, then the source becomes \textit{$\delta$-correlated}, like in e.g. a thermal one following  the fluctuation-dissipation theorem \citep{radiophysics}. The source considered here will have an  Au interface at $z=0$, hence supporting surface plasmons polaritons (SPPs)  in the spectral range under  consideration, and thus enhancing the near field forces \cite{aunon2012photonic}. After performing all the integrations of the form of Eq. (3) for the primary source and for the induced dipoles, (a long but straightforward work, some of whose details are shown in the supplementary information), one sees that $\left\langle E_{i}^{m*}(\mathbf{r}_{0})E_{i}^{m}(\mathbf{r}_{0})\right\rangle =\left\langle H_{i}^{p*}(\mathbf{r}_{0})E_{i}^{p}(\mathbf{r}_{0})\right\rangle =0$. This will be relevant when we discuss the particle Kerker conditions.
\begin{figure}[h]
\begin{centering}
\includegraphics[width=\linewidth]{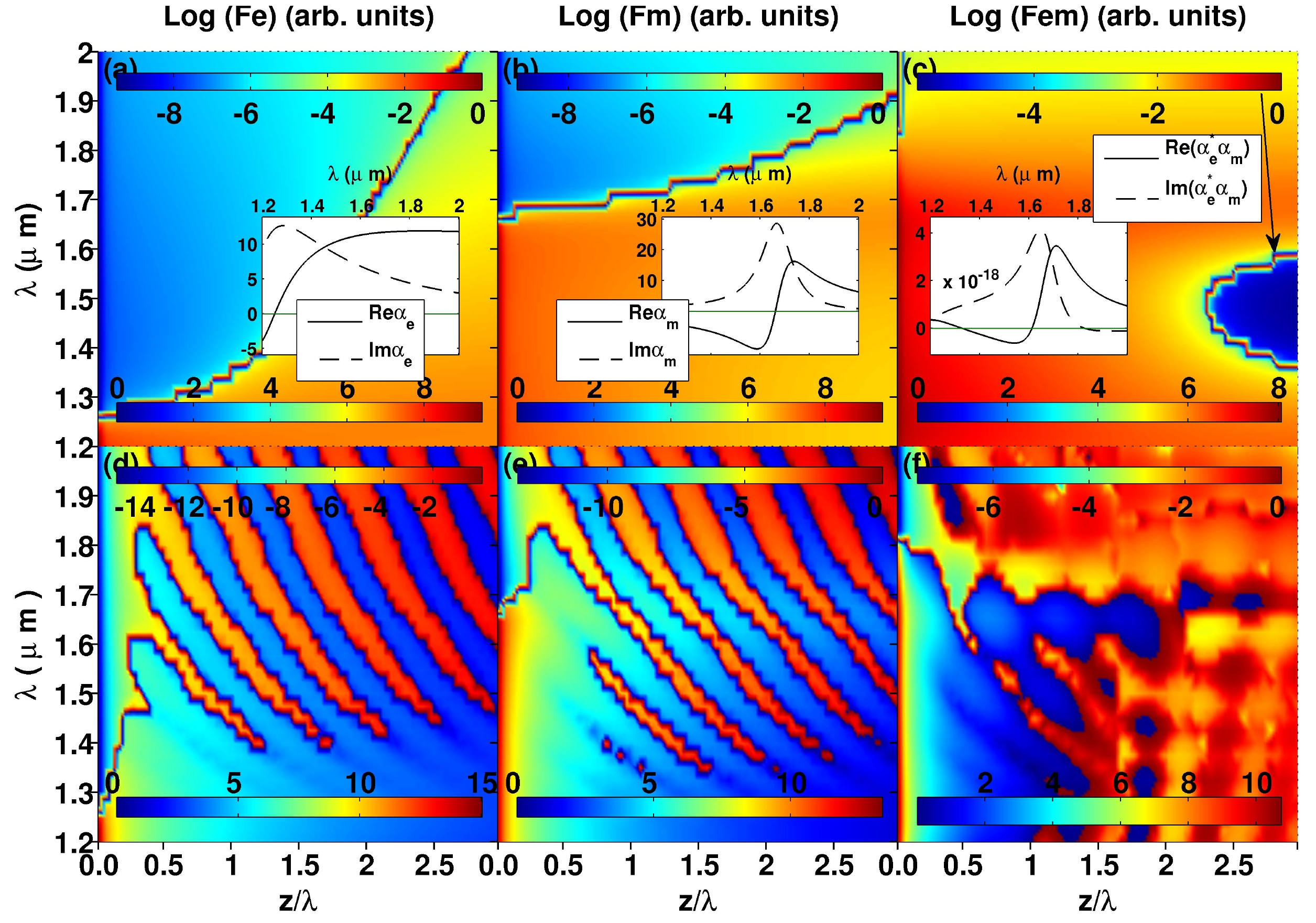}
\par\end{centering}
\caption{(Color online). From left to right: Normalized electric ($F^{e}$), magnetic ($F^{m}$) and interference force ($F^{em}$) from a $\delta$-correlated source. The first horizontal row shows the part of the force due to the field from the fluctuating primary source at $z=0$, [first term in Eq. (2)]. The insets exhibit the behavior of the electric, magnetic and electric-magnetic interaction force components, respectively, normalized to $a^3$. The second horizontal row represents the force from the secondary source constituted by the particle induced dipoles, [second and third terms in Eq. (2)]}
\end{figure}
Fig. 1 shows  the logarithm of the force. This representation aims to clarifying its drastic changes of sign. All the results of this paper will be normalized to the spectrum of the source in order to see the relative weight of each force component. The first horizontal row shows the force $F_{1}$ due to the field impinging from the primary statistical source at $z=0$. The second horizontal row represents the force  $F_{2}$ from the secondary source constituted by the electric and magnetic dipoles induced on the particle. The fluctuating source coherence length $\sigma$ is first assumed to be zero. The inset shows the behavior of the polarizability for the range of wavelengths considered. This helps to understand the color plots. In Figs.  1(a) and 1(b) we  see a line separating the gradient and scattering forces. For a statistically homogeneous source, the gradient force (proportional to $\text{Re}{\alpha_e}\left\langle E_i^*E_i \right\rangle$) is governed uniquely by the evanescent modes and is negative for a particle with $\text{Re}\alpha_e>0$ \citep{aunon2012photonic}, hence, it exponentially decays with the distance z to the source. On the other hand, the scattering force (proportional to $\text{Im}{\alpha_e}\text{Im}{\left\langle E_j^*\partial_iE_j \right\rangle}$)  is positive, i.e. pushing, and constant for any $\mathbf{r}_0$. As  the wavelength grows, $\text{Re}\alpha_e>\text{Im}\alpha_e$, [see the inset in Fig. 1(a)], the gradient force dominates even at  distances larger than  $\lambda$, where  the evanescent modes do not contribute. This is a remarkable new feature of this kind of particles.
Fig. 1 (c) represents the  force component $F_i^{e-m}$ due to interference between the particle induced electric and magnetic dipoles. In the near field this is almost repulsive for any wavelength, however, at distances larger than the wavelength, where the  Poynting vector $\mathbf{S}=\text{Re}(\mathbf{E^*}\times\mathbf{H})/2$ is independent of the distance, we have a zone where this force is negative (the arrow indicates that zone). This kind of action is  known as a \textit{pulling} force \citep{chen2011optical,sukhov2011negative,novitsky2011single}, and its interest has increased in the last years. This last plot shows the relevant role of the magnetodielectric behavior of  these particles in this respect, although in this latter specific case when the two other components: electric and magnetic, are added this pulling effect becomes very small, (the electric-magnetic dipole interference force at $\lambda\simeq1.47\mu m$ and $z\geq2.5\lambda$ is one order of magnitude less than either $F_1^{e}$ or $F_1^{m}$), however by manipulating the fluctuating source, a suitable power of the emitted electromagnetic field is obtained \citep{Novotny2008vdw}  so that  a tractor light field appears, even in the  far zone.

Concerning the force $F_2$ fom the particle induced dipoles, we observe in the second row of Fig. 1 that this force exponentially  decays with the distance z to the primary source plane $z=0$, and its sign depends on that of the particle polarizability; nevertheless,    the oscillating behavior of the Green function due to propagating plane wave components manifests in this force.  We also observe that it is six orders of magnitude larger than its  counterpart  $F_{1}$ from the primary fluctuating source, at least at subwavelength distances $z$. We shall later discuss this.

To get a deeper understanding,  Fig. 2  represents $F_{2}$ for some selected wavelengths and for two different source coherence lengths: $\sigma=0$ and $\sigma=\lambda/4$. For an statistically homogeneous source, the relationship between the electric and magnetic cross-spectral density tensors is \citep{setala2005connection}
\begin{equation}
\varepsilon_{0}\left\langle E_{i}^{*}(\mathbf{r}_{1},\omega)E_{j}(\mathbf{r}_{2})\right\rangle =\mu_{0}\left\langle H_{i}^{*}(\mathbf{r}_{1},\omega)H_{j}^{*}(\mathbf{r}_{2})\right\rangle ,
\end{equation}
hence, in the near field, when the  first or the second Kerker condition holds, one has $F^e_{1}=F^m_{1}$ and $F^e_{1}=-F^m_{1}$, however, in the far zone $F^e_{1}=F^m_{1}$ for any value of $\mathbf{r}_0$. For the Si particle addressed, the Kerker conditions are fullfilled at $\lambda_1\simeq1.825 \mu m$ and $\lambda_2\simeq1.53 \mu m$, (see Figs. 2(a) and 2(b)  and \citep{nietoJOSA011}). We can also see in the inset of Fig. 1(b), that in the range of $\lambda=1.6-1.65 \mu m$ there is a peak in the imaginary part of $\alpha_m$ which predominates  over all other  $\alpha$ parts. The black-dashed-dot line in Fig. 2 represents the force in this peak. In that case,  the magnetic force $F^m_{z,1}$ is one order of magnitude larger than $F^e_{z,1}$ and $F^{em}_{z,1}$, hence the total force on the particle is governed by this magnetic force. This effect, due to the dielectric particle  magnetic  response to the light field, constitutes one of the main results of  this paper.


\begin{figure}[h]
\begin{centering}
\includegraphics[width=\linewidth]{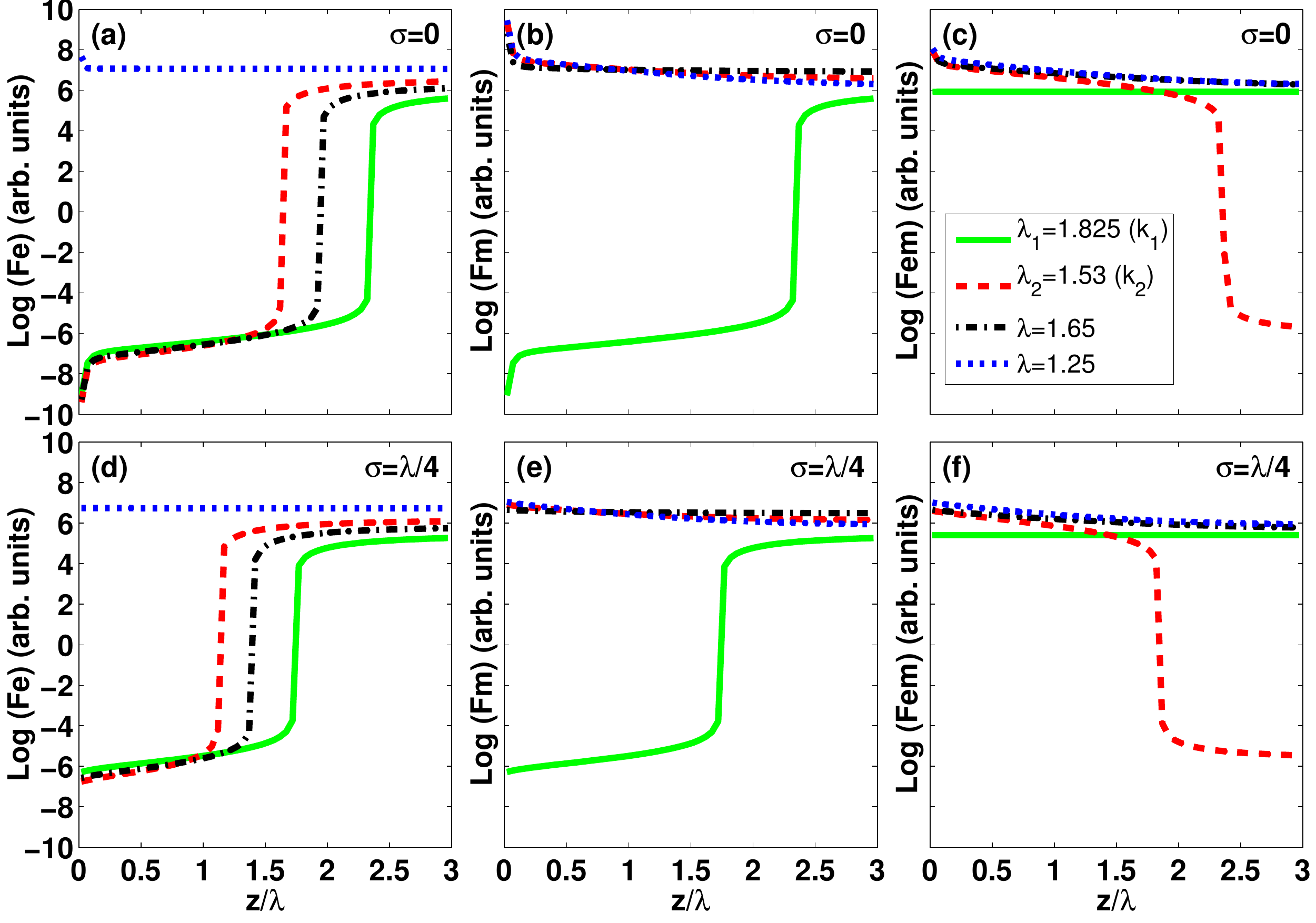} 
\par\end{centering}

\caption{(Color online). Normalized $F_1^{e}$, $F_1^{m}$ and $F_1^{em}$  versus the distance from the plane of the source (in wavelength units) for some values of the wavelength (in $\mu m$). The two Kerker conditions occur at  $\lambda_1$, (K1), and $\lambda_2$, (K2), respectively.}
\end{figure}

We now  address the influence of  the coherence length of the source. This will establish the differences between the mechanical action of partially correlated sources and that from  e.g. thermal sources and blackbodies. The spectral degree of coherence in $k-$space is $\exp [ -  (k \sigma s_{\perp})^2 /2]$, hence, it acts a low-pass filter being  maximum for $\sigma=0$ (i.e. when the source is  $\delta$-correlated). Because of this fact, the evanescent modes present two such filters: the first  is due to the own nature of these evanescent modes while  the second stems from the spatial coherence of the source. The shape of Figs. 2(d)-(f) is  similar to that of Figs. 2(a)-(c), shifted by a  distance $\Delta z \simeq 0.5\lambda$, therefore for $\sigma>\lambda$ the force is solely due to the non-conservative  (scattering) force and to the interference force $F_1^{em}$, which becomes constant and positive or negative depending on the wavelength. It is worth pointing out that the price paid on  increasing the coherence length is expensive, because at the same time there is a reduction of  the force strength by various orders of magnitude [cf. e.g. the forces shown in Figs. 2 (a) and 2 (d)].

\begin{figure}[hbtp]
\begin{centering}
\includegraphics[width=\linewidth]{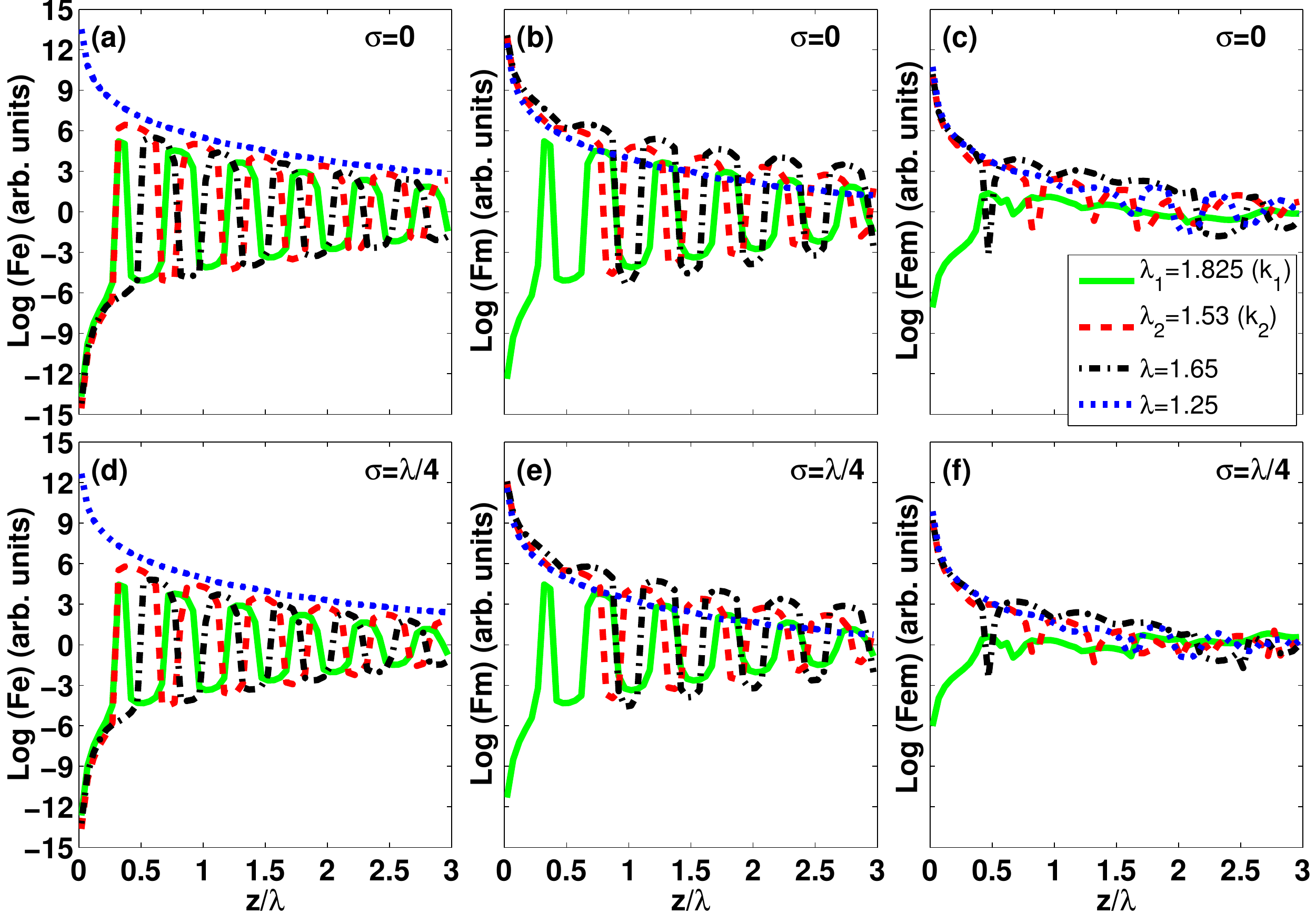}

\caption{Color online). Normalized $F_2^{e}$, $F_2^{m}$ and $F_2^{em}$   versus distance $z$ (in units of wavelength) from the exit plane $z=0$ of the source  for different values of the wavelength (in $\mu m$). The two Kerker conditions are fulfilled at  $\lambda_1$, (K1), and $\lambda_2$, (K2), respectively.}
\par\end{centering}
\end{figure}

Once we understood the role of the coherence length, we turn our study to its influence on the force $F_2$ induced by the secondary source, namely by the particle induced dipoles. Fig. 3 represents $F_{2}$ for the same wavelengths as in Fig. 2. The magnitude of the force in the near-field $z<\lambda$ is much larger than in Fig. 2, thus, the effect of the mechanical action $F_2$ of the field emitted by the particle induced dipoles substantially dominates over that  $F_1$ of the field that is due solely to the stochastic source. Nevertheless, as  the distance $z$ grows, all the fastly oscillating  components, electric, magnetic and that of interference, of this force $F_2$ rapidly tend to zero, and hence is the force $F_1$ from the primary source the one that dominates. As follows from the calculation of $E^{p,m}$ and $H^{p,m}$, the cross spectral density tensor of the electric and magnetic dipoles now are not equal;  therefore, and although at first sight it could seem that similar relationships between the electric and magnetic forces, in Kerker conditions, are fulfilled like for  $F_1$ from the primary source, in fact they are not. 

The role of the coherence length in this case is exactly the same as in Fig. 1;  the magnitude of the force decreases as $\sigma$ grows. Future work should find a minimum value of $\sigma$  for which the Casimir-Polder force predominates over the contributions discussed here.

In summary, we have shown, that the mechanical action on a magnetodielectric small particle from a partially coherent fluctuating source exhibits important new effects at distances shorter than the wavelength. In particular,  the magnetic induced dipole  and its interference with the electric dipole create a landscape of forces  completely different to that previously studied in connection with Van der Waals, Casimir, Liftshitz and rest of radiation forces both in and out of thermodynamic equilibrium. In this respect, Kerker conditions, as a result of the exotic scattering properties of these particles, introduce new relationships in the balance of these new forces over the traditional purely electric forces.   Further experiments should be stimulated by these new effects, including  consequences of Fano resonances.

\section*{Acknowledgements}

The authors acknowledge support from the Spanish Ministerio de Ciencia
e Innovacin (MICINN) through the Consolider NanoLight CSD2007-00046
and FIS2009-13430-C02-01 research grants. J. M. Aun\'{o}n thanks a scholarship
from MICINN.

%

\end{document}